\documentclass[letterpaper, 10 pt, conference]{ieeeconf}
\def\ps@IEEEtitlepagestyle{
  \def\@oddfoot{\mycopyrightnotice}
  \def\@evenfoot{}
}
\def\mycopyrightnotice{
  {\footnotesize IEEE-EMBC-2024, Jul.15-19. Orlando, USA.~\copyright~IEEE All rights reserved. \hfill} 
  \gdef\mycopyrightnotice{}
}
\usepackage{eso-pic}
\newcommand\AtPageUpperMyleft[1]{\AtPageUpperLeft{
 \put(\LenToUnit{0.15\paperwidth},\LenToUnit{-1.5cm}){
     \parbox{0.9\textwidth}{\raggedleft\fontsize{10}{11}\selectfont #1}}
 }}
\newcommand{\conf}[1]{
\AddToShipoutPictureBG*{
\AtPageUpperMyleft{#1}
}
}
\conf{\textit{Accepted Version}: 2024 IEEE / International Conference of the IEEE Engineering in Medicine and Biology Society (EMBC), Jul. 15-19, 2024 - Orlando, USA.
~\textcopyright 2024 IEEE. Personal use of this material is permitted. Permission from IEEE must be obtained for all other uses. More information \href{http://www.ieee.org/publications_standards/publications/rights/index.html}{\textit{here}}}
\newcommand{\placetextbox}[3]{
\setbox0=\hbox{#3}
\AddToShipoutPictureFG{ \put(\LenToUnit{#1\paperwidth},\LenToUnit{#2\paperheight}){\vtop{{\null}\makebox[0pt][c]{#3}}}}
}
\placetextbox{.33}{0.05}{~\copyright~IEEE All rights reserved. IEEE-EMBC-2024, Jul.15-19. Orlando, USA.\hfill}
\IEEEoverridecommandlockouts  
\usepackage{graphicx}
\usepackage[nointegrals]{wasysym}
\usepackage{amsmath}
\usepackage{tabularx}
\usepackage{comment}
\usepackage{xcolor}
\usepackage{subcaption}
\usepackage{float}
\usepackage{gensymb}
\usepackage{textcomp}
\usepackage{multirow}
\usepackage{booktabs}
\usepackage{hyperref}

\title{\LARGE \bf
Validation of the estimated Effect of Ankle Foot Orthoses on Spinal Cord Injury Gait Using Subject-Adjusted Musculoskeletal Models
}

\author{Sergio Galindo-Leon$^{1}$, Inge Eriks-Hoogland$^{2}$, Kenji Suzuki $^{3}$ and Diego Paez-Granados$^{4}$
\thanks{This work was supported in part by the Schweizer Paraplegiker Stiftung (SPS),  the ETH Zürich Foundation ETH-SPS Digital Transformation in Personalized Health Care for SCI, and the JSPS - ETHZ Leading House Asia's Young Researchers Exchange Programme – Special 2023}
\thanks{$^{1}$Sergio Galindo is with the School of Integrative and Global Majors, University of Tsukuba, Tsukuba, Japan and with the SCAI Lab, ETH Zurich, Switzerland
        {\tt\small sergio@ai.iit.tsukuba.ac.jp, sergioalbert.galindoleon@hest.ethz.ch}}%
\thanks{$^{2}$Inge Erkis-Hoogland is with the Swiss Paraplegic Center (SPC), Switzerland
        {\tt\small inge.eriks@paraplegie.ch}}
\thanks{$^{3}$Kenji Suzuki is with the Faculty of Systems, Information and Engineering, University of Tsukuba, Tsukuba, Japan
        {\tt\small kenji@ieee.org}}
\thanks{$^{4}$Diego Paez-Granados is with the SCAI Lab, ETH Zurich, Swiss Paraplegic Research (SPF), Switzerland
        {\tt\small diego.paez@hest.ethz.ch}}
}

\begin{document}
\maketitle
\thispagestyle{empty}
\pagestyle{empty}

\begin{abstract}
Simulation of assistive devices on pathological gait through musculoskeletal models offers the potential and advantages of estimating the effect of the device in several biomechanical variables and the device characteristics ahead of manufacturing. In this study, we introduce a novel musculoskeletal modelling approach to simulate the biomechanical impact of ankle foot orthoses (AFO) on gait in individuals with spinal cord injury (SCI). Leveraging data from the Swiss Paraplegic Center, we constructed anatomically and muscularly scaled models for SCI-AFO users, aiming to predict changes in gait kinematics and kinetics. The importance of this work lies in its potential to enhance rehabilitation strategies and improve quality of life by enabling the pre-manufacturing assessment of assistive devices. Despite the application of musculoskeletal models in simulating walking aids effects in other conditions,
no predictive model currently exists for SCI gait. Evaluation through RMSE showed similar results compared with other pathologies, simulation errors ranged between 0.23 to 2.3 degrees in kinematics. Moreover, the model was able to capture ankle joint muscular asymmetries and predict symmetry improvements with AFO use. However, the simulation did not reveal all the AFO effects, indicating a need for more personalized model parameters and optimized muscle activation to fully replicate orthosis effects on SCI gait.

\end{abstract}


\section{INTRODUCTION}
The simulation of pathological and assisted gait through musculoskeletal models has been used to support the development and evaluation of a wide range of devices including prosthesis, exoskeletons, and orthosis. 
This process aims to model and estimate biomechanical and physiological responses even before the prototyping stage \cite{Ackermann2010PredictiveRehabilitation}, such processes have been proposed for the design of exoskeletons and aids \cite{Paez-Granados2022PersonalTransfer,Aftabi2021Simulation-basedRunning}, where exoskeletons could be tailor-made using a biomechanics model and forward dynamic simulation. However, the complexity of pathological gait still requires investigating parametrization on both humans and aids. 
In particular, ankle foot orthosis (AFO), which restricts ankle motion is widely used for a great variety of clinical applications including muscular disorders, fractures and neuromuscular pathologies such as cerebral palsy, stroke and spinal cord injury (SCI) \cite{Cikajlo2016TheImpairments,Everaert2023TheStudy}. 

Orthoses also play an important role in the rehabilitation and prognosis of ambulatory SCI individuals. AFOs provide weight support, paralyzed muscle compensation, posture correction and also help in motor dysfunction, which ultimately affects the activities of daily living and quality of life of the users\cite{Cui2023Advances2013}. The biomechanical effects of AFO in SCI individuals include an increase in step length and walking capacity, and also, have been shown to help in the prevention of lower limb deformities\cite{Cui2023Advances2013}. Moreover, predicting the effects of an AFO on gait could also be used for a more accurate design and prescription of gait aids as the effects of orthotic devices are generally individual-specific \cite{Cui2023Advances2013}. Therefore, several attempts to replicate the impacts of AFO in pathological gait have been performed, but most advanced simulations still present only healthy subject gait analysis and effect \cite{Auer2022BiomechanicalSystem}. 

\begin{figure*}
  \includegraphics[width=\linewidth]{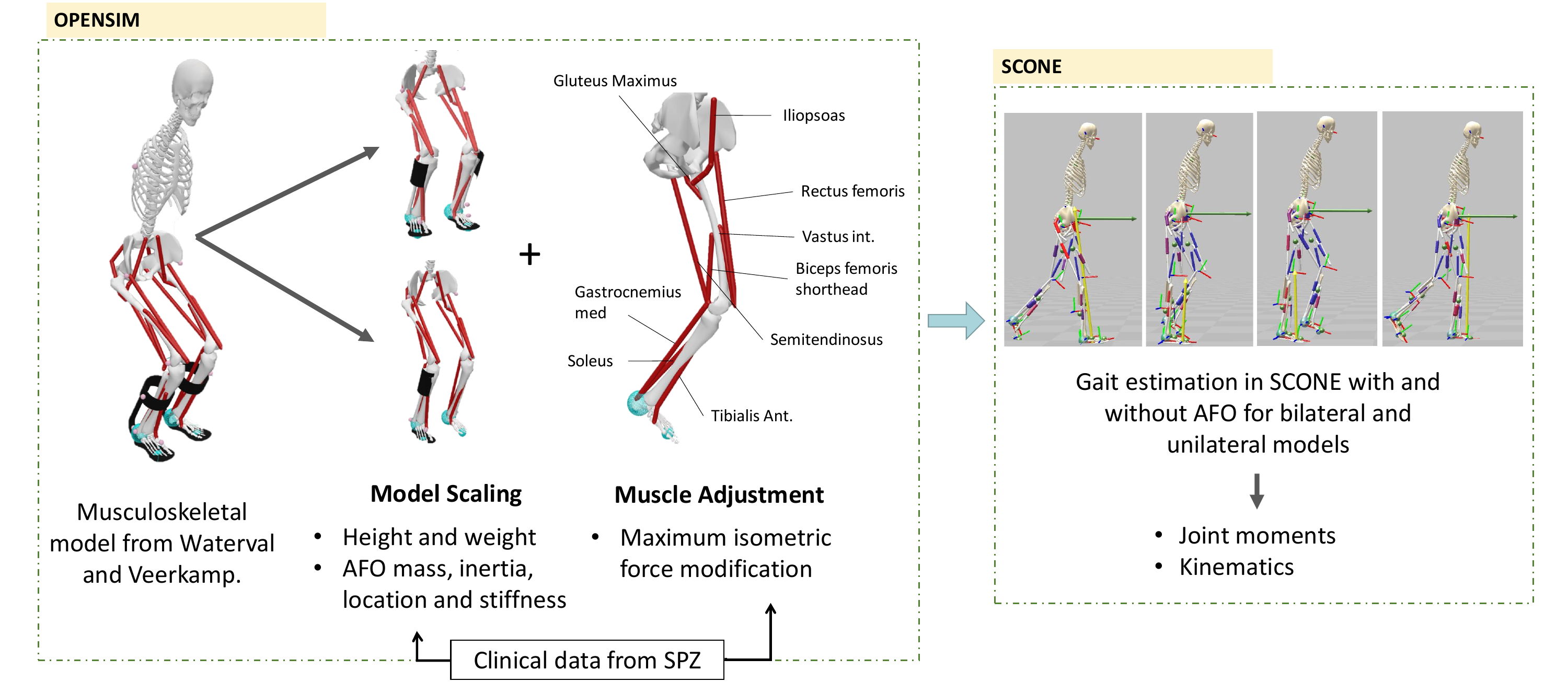}
  \caption{Gait simulation pipeline for including unilateral and bilateral effects of ankle-foot orthosis on the kinematics and kinetics of spinal cord injury individuals.}
  \label{fig:methods}
\end{figure*}

For the simulation of AFO gait, most models still required previous acquisition of kinematics and ground reaction forces to compute the remaining biomechanical variables as moments and powers\cite{Rosenberg2017SimulatedGait, Rosenberg2017SimulatedGait, Yamamoto2019EffectModel}. Even though these simulations were able to elucidate some of the effects of the AFO they were not predictive as they lacked a forward simulation of gait and required the input of experimentally acquired AFO kinematics. Because of this, Kiss, et al (2021)\cite{Kiss2021PredictingStudy} introduced the first predictive study of the effects of an AFO in pathological gait, specifically, in bilateral calf muscle weakness. Using a reflex-based controller and a close loop optimization of the metabolic cost of transport, they were able to simulate the effects of different AFO stiffness in the kinematic, kinetic, and muscular trends of individuals. These results were further validated by Waterval, et al. with retrospective data from AFO users \cite{Waterval2022TheStudy,Waterval2023InteractingStudy}. 

Even though musculoskeletal models have been used for simulating and estimating AFO gait in cerebral palsy, crouch gait and bilateral plantar-flexor weakness\cite{Rosenberg2017SimulatedGait, Rosenberg2017SimulatedGait, Yamamoto2019EffectModel,Kiss2021PredictingStudy,Waterval2022TheStudy}, there is still not any predictive musculoskeletal model for SCI assisted gait that can be used to estimate these effects. 

In this work, we proposed and evaluated gait simulation models for individuals with spinal cord injury (SCI), offering three main contributions: (1) a bilateral SCI musculoskeletal model that incorporates specific anatomical and muscular strength characteristics of SCI subjects, based on a calf muscle weakness framework. (2) a method for parametrizing ankle foot orthoses (AFOs) that supports both unilateral and bilateral use cases. (3) Validation of our gait simulation model using data from SCI subjects, with and without AFO use, by comparing kinematic and kinetic data to actual measurements.
These contributions aim to enhance the precision of rehabilitation and assistive device optimization by providing a more accurate representation of the biomechanical impact of AFOs in SCI patients.

\section{METHODOLOGY}
To produce and validate unilateral AFO spinal cord injury (SCI) gait simulations with the available data from the Swiss Paraplegic Center, we developed a bilateral plantar flexor weakness musculoskeletal model based on (Waterval and Veerkamp) \cite{Waterval2022TheStudy, Waterval2021IndividualRecommendations, Veerkamp2021EvaluatingGait, Kiss2022MinimizationWeakness} along with their gait state-dependent reflex base controller\cite{Waterval2021ValidationGait} in SCONE\cite{Geijtenbeek2019SCONE:Motion} to obtain full gait cycles of both AFO assisted and unassisted SCI gait. 
We generated personalized musculoskeletal models for each of the subjects in the dataset in OpenSim\cite{Delp2007OpenSim:Movement}, based on their anatomical characteristics and muscular state assessment by adjusting the bilateral calve weakness model accordingly. This model contains nine hill-type muscles that follow the Millard Equilibrium model per lower limb. The muscles included correspond to Tibialis Anterior, Soleus, Gastrocnemius medialis, Vastus intermedius, Rectus Femoris, Semitendinosus, Biceps Femoris short head, Gluteus Maximus and Iliopsoas \cite{Waterval2021ValidationGait}. The muscle path, optimal fiber length, and tendon slack length were also maintained as described in the work of Waterval\cite{Waterval2021ValidationGait}. Regarding the contact forces with the ground this model uses two viscoelastic Hunt-Crossley contact spheres on each foot, located in the calcaneus and the forefoot\cite{Waterval2021ValidationGait}
Following this, we generated forward gait simulations in SCONE\cite{Geijtenbeek2019SCONE:Motion} for each of the models and evaluated them using the RMSE for the ankle, knee and hip kinematics and moments. A summary of the methodology is shown in figure \ref{fig:methods}.

\subsection{Dataset}
The dataset collected at the Swiss Paraplegic Center \textit{(Gait in ambulatory individuals with spinal cord injury, ethical approval: EKNZ-2022-00935)}, corresponds to the gait of 5 subjects with spinal cord injury at levels L2 or below who can ambulate on their own with the use of unilateral (n=3. left=1, AFO: fior-gentz neuroswing; right = 2, AFO: Allard Bluerocker \& Allard Toe-off) or bilateral (n=2, AFO: Allard Toe-off) ankle foot orthosis support. The data includes three-dimensional kinematics of the hip, knee and ankle over several steps segmented from consecutive left and right heel strikes. 
Additionally, the dataset contains the height and weight of the subject as well as the assessment of the muscular state performed by a physician and evaluated under the Medical Research Council Manual Muscle Testing scale. 
Here, the muscular strengths of the hip flexors, knee extensors, ankle dorsiflexion, ankle plantar flexors and long toe extensors were evaluated in the [0-5] range being 0 no muscle activation and 5 Muscle activation against examiner’s full resistance under the full range of motion, data is summarized in table \ref{tab:subject_muscles}.

\begin{table*}[]
\centering
\caption{Clinical, anatomical and muscular details of the dataset used for the musculoskeletal model adjustment and forward gait simulations of bilateral and unilateral ACI-AFO gait}
\label{tab:subject_muscles}
\resizebox{\textwidth}{!}{%
\begin{tabular}{@{}llllcccccccccccccccccc@{}}
\toprule
\multirow{2}{*}{Type} & \multirow{2}{*}{\begin{tabular}[c]{@{}l@{}}individual\\  ID\end{tabular}} & \multirow{2}{*}{\begin{tabular}[c]{@{}l@{}}Height \&\\ weight\end{tabular}} & \multirow{2}{*}{\begin{tabular}[c]{@{}l@{}}Muscular\\ Evaluation\end{tabular}} & \multicolumn{2}{c}{Hamstrings} & \multicolumn{2}{c}{\begin{tabular}[c]{@{}c@{}}Biceps\\ femoris\\ (Short H)\end{tabular}} & \multicolumn{2}{c}{\begin{tabular}[c]{@{}c@{}}Gluteus \\ maximus\end{tabular}} & \multicolumn{2}{c}{Iliopsoas} & \multicolumn{2}{c}{\begin{tabular}[c]{@{}c@{}}Rectus\\ femoris\end{tabular}} & \multicolumn{2}{c}{\begin{tabular}[c]{@{}c@{}}Vastus\\ interm\end{tabular}} & \multicolumn{2}{c}{Gastroc} & \multicolumn{2}{c}{Soleus} & \multicolumn{2}{c}{\begin{tabular}[c]{@{}c@{}}Tibialis \\ anterior\end{tabular}} \\ \cmidrule(l){5-22} 
 &  &  &  & L & R & L & R & L & R & L & R & L & R & L & R & L & R & L & R & L & R \\ \cmidrule(r){1-4}
\multirow{2}{*}{Bilateral} & 1 & \begin{tabular}[c]{@{}l@{}}172 cm\\ 64 kg\end{tabular} & \begin{tabular}[c]{@{}l@{}}Muscle\\ strength \\ {[}0.0-1.0{]}\end{tabular} & 3/5 & 3/5 & 3/5 & 3/5 & 3/5 & 3/5 & 3/5 & 3/5 & 3/5 & 4/5 & 3/5 & 2/5 & 2.5/5 & 1.5/5 & 2.5/5 & 1.5/5 & 3/5 & 0.5/5 \\
 & 2 & \begin{tabular}[c]{@{}l@{}}165cm\\ 53 kg\end{tabular} & \begin{tabular}[c]{@{}l@{}}Muscle\\ strength \\ {[}0.0-1.0{]}\end{tabular} & 4/5 & 3/5 & 5/5 & 4/5 & 5/5 & 4/5 & 4/5 & 4/5 & 4/5 & 4/5 & 5/5 & 4/5 & 5/5 & 4/5 & 5/5 & 4/5 & 3/5 & 4/5 \\
\multirow{2}{*}{Right} & 3 & \begin{tabular}[c]{@{}l@{}}176 cm\\ 77 kg\end{tabular} & \begin{tabular}[c]{@{}l@{}}Muscle\\ strength \\ {[}0.0-1.0{]}\end{tabular} & 5/5 & 3/5 & 5/5 & 3/5 & 5/5 & 2/5 & 4/5 & 2.5/5 & 5/5 & 2.5/5 & 3/5 & 2.5/5 & 5/5 & 2/5 & 5/5 & 2/5 & 5/5 & 1/5 \\
 & 4 & \begin{tabular}[c]{@{}l@{}}179 cm\\ 85kg\end{tabular} & \begin{tabular}[c]{@{}l@{}}Muscle\\ strength \\ {[}0.0-1.0{]}\end{tabular} & 5/5 & 3/5 & 5/5 & 5/5 & 5/5 & 5/5 & 4/5 & 2/5 & 4/5 & 3/5 & 5/5 & 5/5 & 5/5 & 4/5 & 5/5 & 4/5 & 2/5 & 3/5 \\
Left & 5& \begin{tabular}[c]{@{}l@{}}173 cm\\ 72kg\end{tabular} & \begin{tabular}[c]{@{}l@{}}Muscle\\ strength \\ {[}0.0-1.0{]}\end{tabular} & 3/5 & 4/5 & 3/5 & 4/5 & 4/5 & 4/5 & 4/5 & 4/5 & 4.5/5 & 4.5/5 & 4/5 & 4/5 & 5/5 & 4/5 & 5/5 & 4/5 & 3/5 & 4.5/5 \\ \bottomrule
\end{tabular}%
}
\end{table*}
\subsection{Spinal Cord Injury Model}

To generate the spinal cord injury - ankle foot orthosis (SCI-AFO) models we modified and used the seven segments, armless, nine degrees of freedom bilateral plantar flexor weakness model of Waterval and Veerkamp \cite{Waterval2022TheStudy,Waterval2021IndividualRecommendations,Veerkamp2021EvaluatingGait,Kiss2022MinimizationWeakness}. 
First, we scaled the model to the height and weight of each of the subjects using the scale tool in OpenSim\cite{Delp2007OpenSim:Movement}. Then, to mimic the muscular state of the subjects, we multiplied the maximum isometric force of each of the model's muscle actuators by the score of the corresponding muscle or group of muscles in the Medical Research Council Manual Muscle Testing scale, normalized by the maximum possible score (5) (table \ref{tab:subject_muscles}). 

For the AFOs, we created an AFO shank and footplate and replaced the existing geometries and locations of these bodies in conformity with the AFO worn by the subjects either unilaterally or bilaterally. As the AFO locations and orientations changed from the original model to a ventral position, we recalculated the centres of mass and moments of inertia for the new geometries and placements. For the subjects using Allard Toe-off and Allard Bluerocker AFOs the total mass of the shank and footplate was set to 170g\cite{Jackson2021TheStudy} (shank 70g and footplate 100g) and the stiffness, represented by a coordinate limit force spring damper object in the ankle joint, was set to 2.02 Nm/deg and 3.5 Nm/deg respectively\cite{Shuman2022MultiplanarOrthoses}. For the subject using the fior-gentz Neuroswing AFO the mass of the device was set to 450g (250g shank and 200g for the footplate) and the 
stiffness to 2.2 Nm/deg\cite{Ploeger2019StiffnessStudy}. For the bilateral cases the stiffness was maintained equal for both limbs at 2.2 Nm/deg as the users were prescribed Allard Toe-off in both limbs. Lastly, for the unassisted simulations, we removed the shank and footplates from the models and set the stiffness to 0.01 Nm/deg. Increasing the number of parameters in the simulations by including the muscular strength, the orientation and, the mass of the AFO, not only allows for a better representation of the subjects but also, corresponds to a novel pilot assessment for pathological gait simulations with orthotic devices 

\subsection{Forward gait Simulations and validation}
We generated forward gait simulations in SCONE\cite{Geijtenbeek2019SCONE:Motion} for each of the models and cases (AFO and Non AFO gait)
through minimization the cost function developed by Veerkamp, et al. and Waterval, et al. \cite{Veerkamp2021EvaluatingGait,Waterval2021ValidationGait} using the Covariance Matrix Adaptation Evolution Strategy (CMA-ES).

The optimization was sustained until the variation in the cost function between generations was smaller than 0.1\% or the number of iterations reached the maximum, set in 1200. The obtained ankle, knee and hip kinematics and moments were plotted (Figures \ref{fig:bilateral} and \ref{fig:unilateral}) and compared against the data from the Swiss Paraplegic Center dataset using the root mean square error (RMSE), normalized to the standard deviation of the data of each of the subjects \cite{Waterval2021ValidationGait}. 

\section{RESULTS}

\subsection{Bilateral kinematics}

The models and simulations for  bilateral unassisted gait can represent and show the subjects' asymmetries regarding the muscular force state within their limbs, shown in table \ref{tab:subject_muscles}. In the ankle joint, the simulated kinematics show an increased range of motion (ROM) in the left ankle for both subjects, following the disproportion in their ankle flexor strengths. Similarly, the knee kinematics show the same increase in the ROM for subject 4 but not for subject 5, which also responds to their corresponding muscular states of the rectus femoris and vastus intermedius. However, these asymmetries are not shown in the kinematics of the individuals, differing from the simulated ones in terms of profile and range of motion in the ankle joint and also, in the knee joint for one of the subjects. Nevertheless, the RMSE values (table \ref{tab:kinematics}) present values similar to the ones obtained in previous validations for the knee and hip (1.13, 2.63) \cite{Waterval2021ValidationGait}, while being higher for the ankle joint. 

\begin{figure*}[tb]
\begin{subfigure}{\textwidth}
    \centering
    \includegraphics[width=1\textwidth]{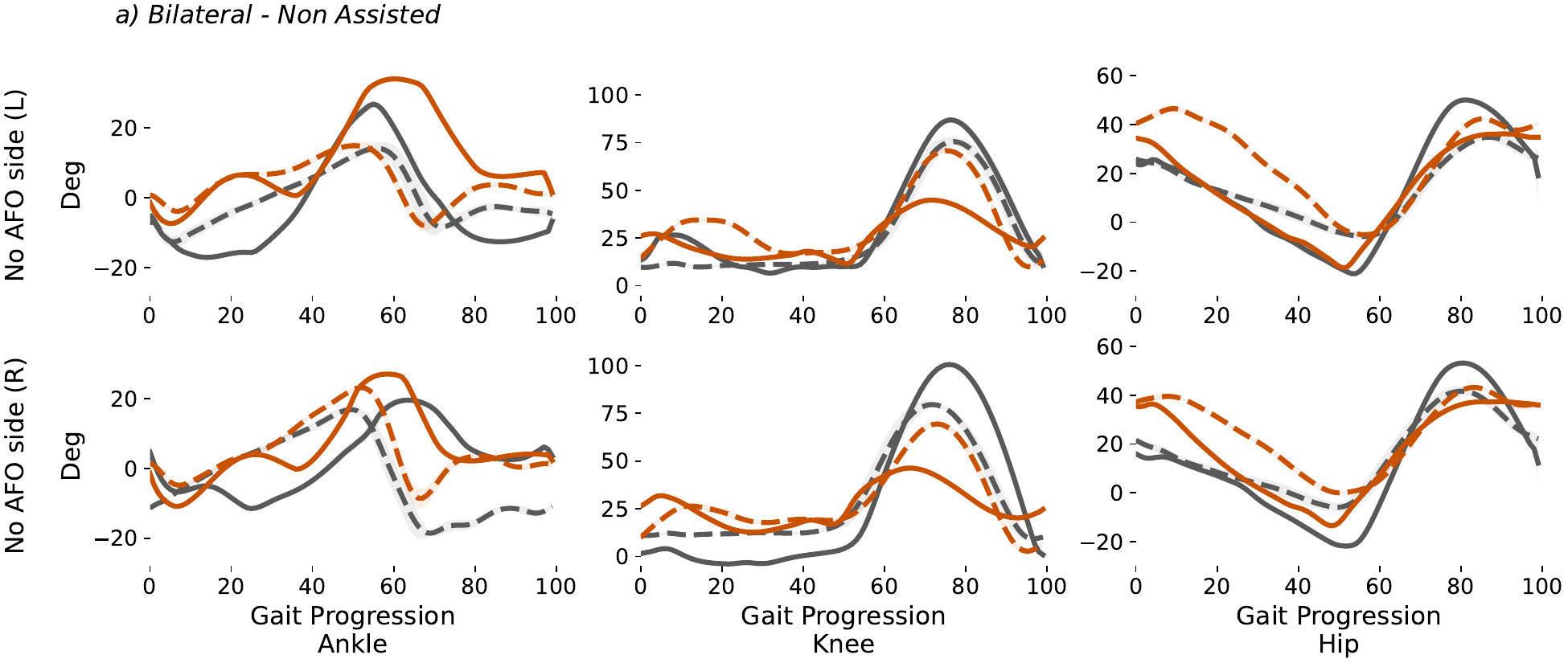}
    \label{fig:second}\\~\\
\end{subfigure}
\begin{subfigure}{\textwidth}
    \centering
    \includegraphics[width=1\textwidth]{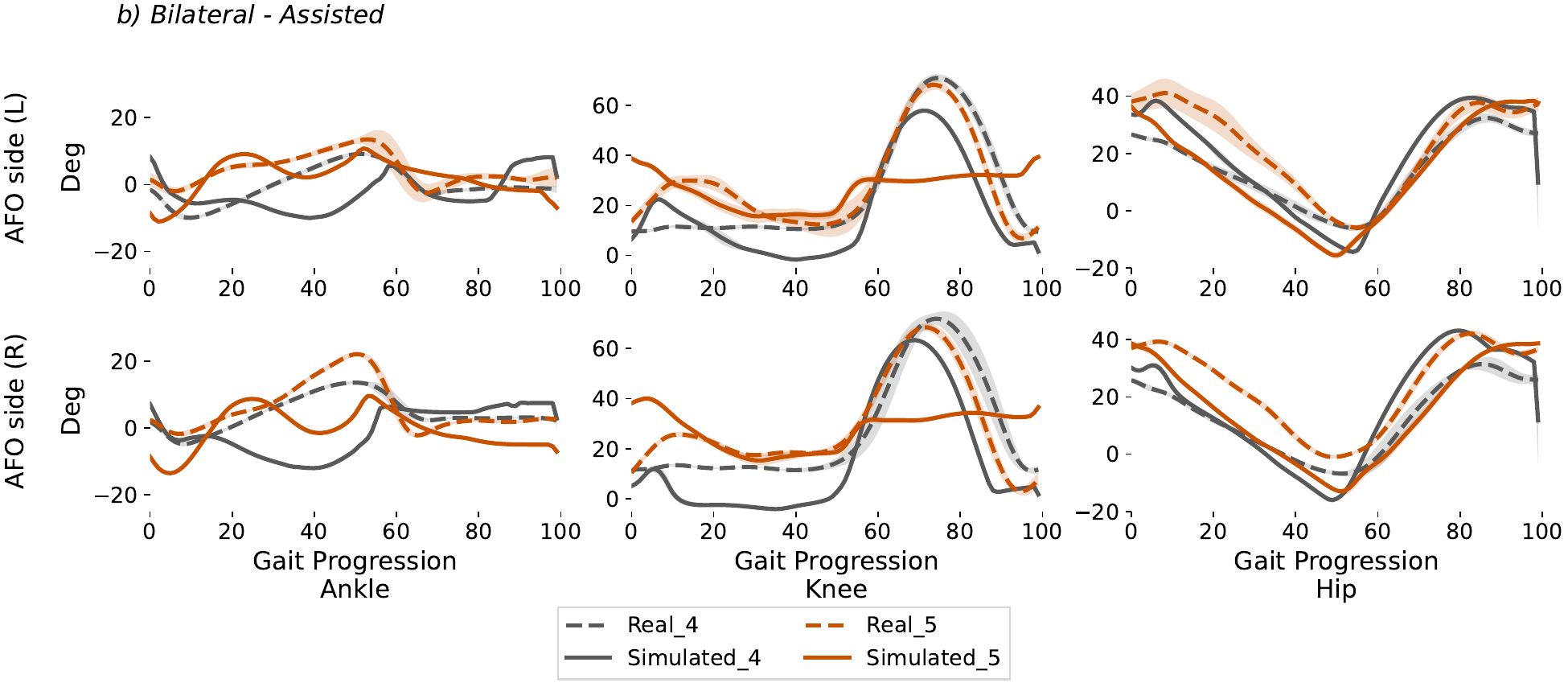}
    \label{fig:bilaterals}
\end{subfigure}
\caption{Obtained kinematics of SCI-AFO and SCI-Non AFO walk from the forward gait simulations and clinical data for the Bilateral AFO models and users.}
\label{fig:bilateral}
\end{figure*}

In comparison to the unassisted case, the bilaterally assisted model can generate more symmetric kinematics in the ankle joint despite the muscular strength disproportions in the subjects, evidencing the effect of the ankle torque input from the AFO in the user's predicted kinematics as seen in figure \ref{fig:bilateral}. Still, apart from the symmetry improvement and ROM adjustment within limbs, the simulated kinematic changes caused by the AFO do not strongly correspond to the effects observed by the use of an AFO in the individual in the ankle and knee joint. In the hip, on the contrary, there is a decrease in the RMSE showing an improvement in the kinematic prediction for this joint in comparison to the unassisted case. Lastly, the RMSE  for the assisted bilateral simulations were found to be generally increased compared with the unassisted case while still being comparable to previous validations \cite{Waterval2021ValidationGait}. 

\begin{figure*}[htbp]
\begin{subfigure}{\textwidth}
    \centering
    \includegraphics[width=1\textwidth]{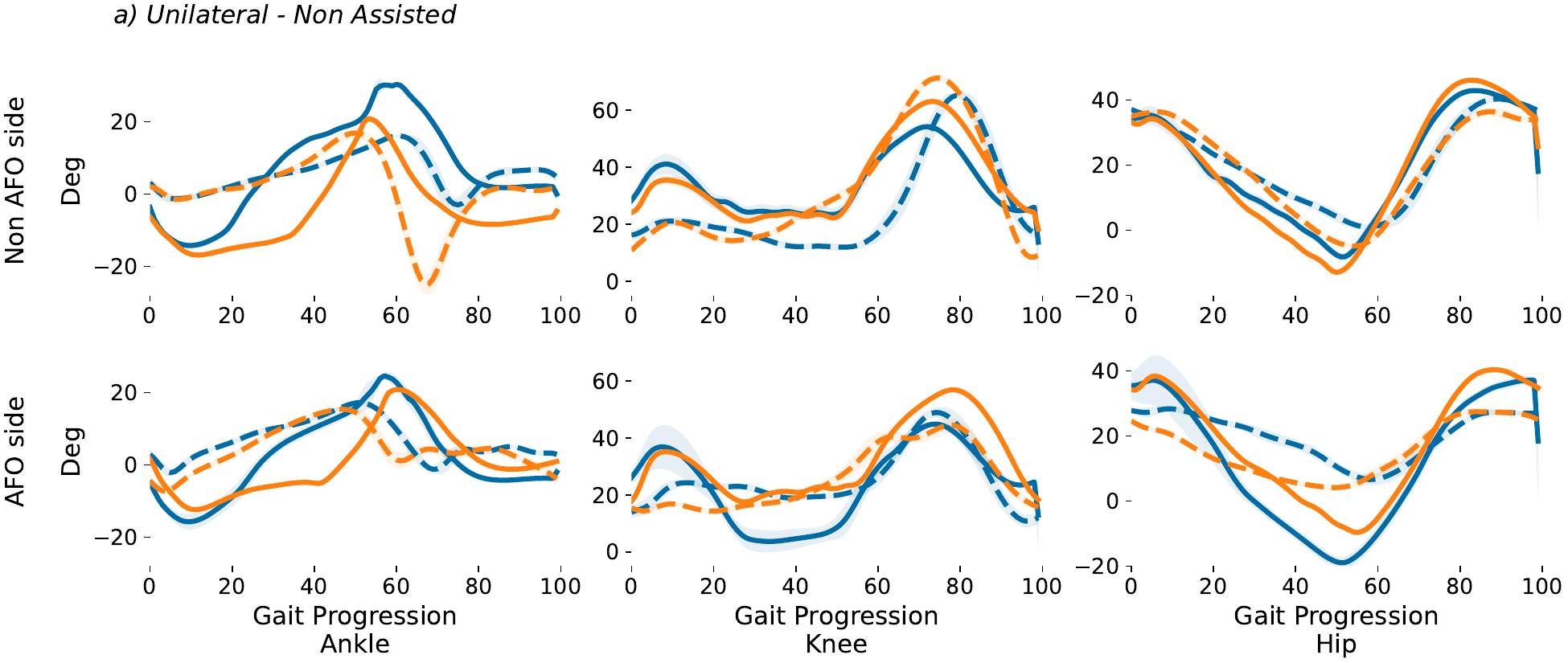}
    \label{fig:seconds}\\~\\
\end{subfigure}
\begin{subfigure}{\textwidth}
    \centering
    \includegraphics[width=1\textwidth]{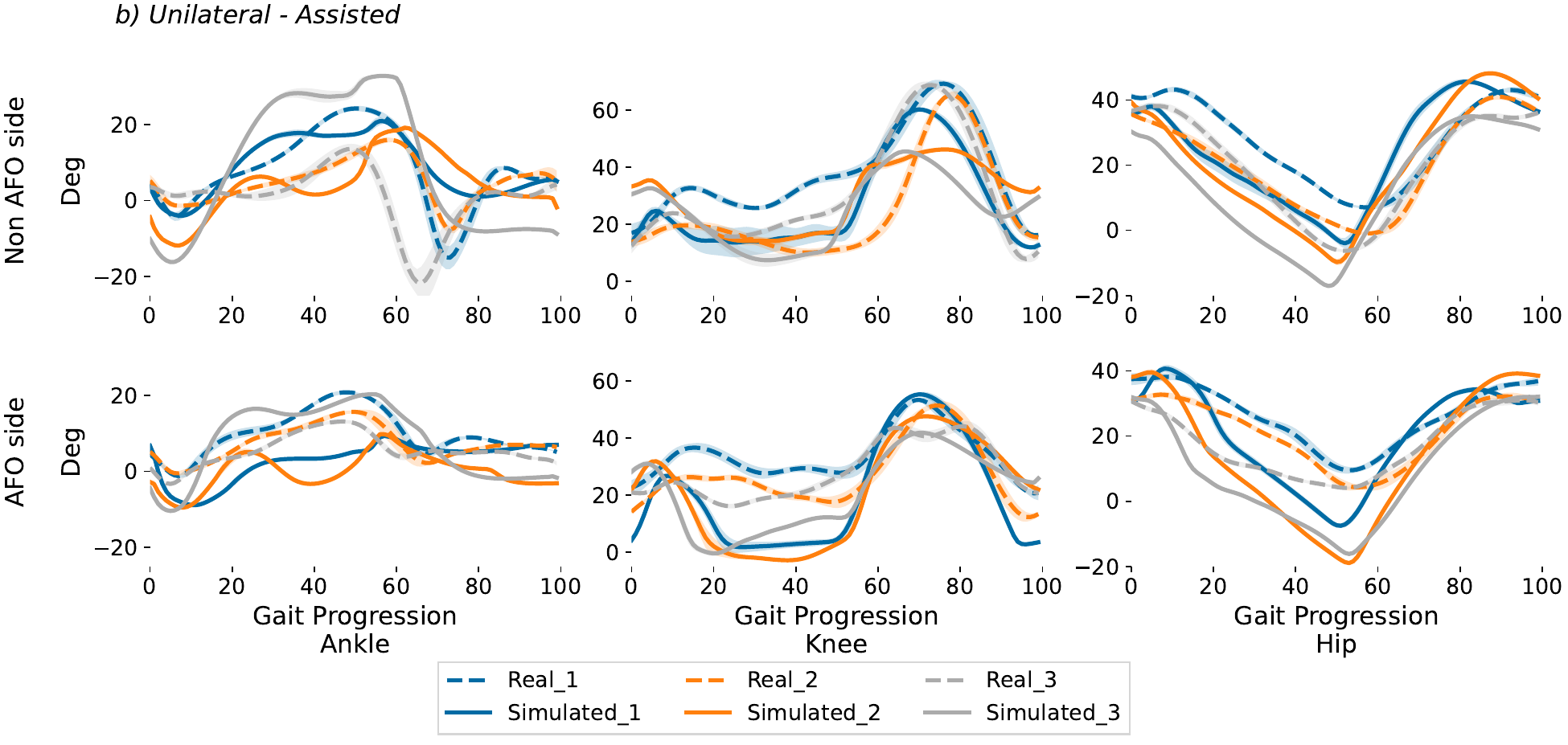}
    \label{fig:unilaterals}
\end{subfigure}
        
\caption{Obtained kinematics of SCI-AFO and SCI-Non AFO walk from the forward gait simulations and clinical data for the unilateral AFO models and users.}
\label{fig:unilateral}
\end{figure*}

\begin{table*}[htbp]
\centering
\caption{RMSE values for the kinematics of the gait simulations with and without AFO use}
\label{tab:kinematics}
\begin{tabular}{@{}llllllllllllll@{}}
\toprule
\multicolumn{4}{l}{\textit{\textbf{Unilateral Case}}} & \multicolumn{3}{l}{RMSE} &  & \multicolumn{4}{l}{\textit{\textbf{Bilateral Case}}} & \multicolumn{2}{l}{RMSE} \\ \midrule
\multicolumn{4}{l}{Subject ID} & 1 & 2 & 3 &  & \multicolumn{4}{l}{Subject ID} & 4 & 5 \\\midrule
\multirow{6}{*}{\begin{tabular}[c]{@{}l@{}}AFO\\ Supported\\ Gait\end{tabular}} & \multirow{3}{*}{\begin{tabular}[c]{@{}l@{}}AFO\\ side\end{tabular}} & \multirow{3}{*}{\begin{tabular}[c]{@{}l@{}}Angle\\ {[}Deg{]}\end{tabular}} & Ankle & 1.55 & 1.97 & 1.76 &  & \multirow{6}{*}{\begin{tabular}[c]{@{}l@{}}AFO\\ Supported\\ Gait\end{tabular}} & \multirow{3}{*}{\begin{tabular}[c]{@{}l@{}}AFO side\\ (R)\end{tabular}} & \multirow{3}{*}{\begin{tabular}[c]{@{}l@{}}Angle\\ {[}Deg{]}\end{tabular}} & Ankle & 2.21 & 1.30 \\
 &  &  & Knee & 2.08 & 1.36 & 1.08 &  &  &  &  & Knee & 0.75 & 0.97 \\
 &  &  & Hip & 1.06 & 1.39 & 1.06 &  &  &  &  & Hip & 0.70 & 0.71 \\
 & \multirow{3}{*}{\begin{tabular}[c]{@{}l@{}}Non AFO\\ side\end{tabular}} & \multirow{3}{*}{\begin{tabular}[c]{@{}l@{}}Angle\\ {[}Deg{]}\end{tabular}} & Ankle & 0.60 & 1.29 & 2.44 &  &  & \multirow{3}{*}{\begin{tabular}[c]{@{}l@{}}AFO side\\ (L)\end{tabular}} & \multirow{3}{*}{\begin{tabular}[c]{@{}l@{}}Angle\\ {[}Deg{]}\end{tabular}} & Ankle & 1.48 & 1.04 \\
 &  &  & Knee & 0.85 & 0.73 & 0.76 &  &  &  &  & Knee & 0.58 & 0.95 \\
 &  &  & Hip & 0.89 & 0.58 & 0.73 &  &  &  &  & Hip & 0.63 & 0.73 \\
 &  &  &  &  &  &  &  &  &  &  &  &  &  \\
\multirow{6}{*}{\begin{tabular}[c]{@{}l@{}}Non\\ Supported\\ Gait\end{tabular}} & \multirow{3}{*}{\begin{tabular}[c]{@{}l@{}}AFO\\ side\end{tabular}} & \multirow{3}{*}{\begin{tabular}[c]{@{}l@{}}Angle\\ {[}Deg{]}\end{tabular}} & Ankle & - & 1.82 & 1.87 &  & \multirow{6}{*}{\begin{tabular}[c]{@{}l@{}}Non\\ Supported\\ Gait\end{tabular}} & \multirow{3}{*}{\begin{tabular}[c]{@{}l@{}}AFO side\\ (R)\end{tabular}} & \multirow{3}{*}{\begin{tabular}[c]{@{}l@{}}Angle\\ {[}Deg{]}\end{tabular}} & Ankle & 1.59 & 1.14 \\
 &  &  & Knee & - & 1.04 & 0.93 &  &  &  &  & Knee & 0.69 & 0.69 \\
 &  &  & Hip & - & 2.27 & 1.22 &  &  &  &  & Hip & 0.62 & 0.72 \\
 & \multirow{3}{*}{\begin{tabular}[c]{@{}l@{}}Non AFO\\ side\end{tabular}} & \multirow{3}{*}{\begin{tabular}[c]{@{}l@{}}Angle\\ {[}Deg{]}\end{tabular}} & Ankle & - & 1.85 & 1.43 &  &  & \multirow{3}{*}{\begin{tabular}[c]{@{}l@{}}AFO side\\ (L)\end{tabular}} & \multirow{3}{*}{\begin{tabular}[c]{@{}l@{}}Angle\\ {[}Deg{]}\end{tabular}} & Ankle & 1.14 & 2.45 \\
 &  &  & Knee & - & 0.92 & 0.47 &  &  &  &  & Knee & 0.34 & 0.75 \\
 &  &  & Hip & - & 0.55 & 0.59 &  &  &  &  & Hip & 0.80 & 0.93 \\ \cmidrule(lr){4-8} \cmidrule(l){12-14} 
\end{tabular}
\end{table*}

\subsection{Unilateral kinematics}
The simulated kinematics for the unilateral models are shown in figure \ref{fig:unilateral}. The unilateral unassisted simulations show more consistent kinematic predictions compared to the bilateral case despite an overall increase in the RMSE (table \ref{tab:kinematics}). In the ankle joint, it predicts an initial plantarflexion after heel strike and the later dorsiflexion peak which is also observed in the unassisted gait of the subjects.  Similarly, the simulated kinematics exhibit both of the knee flexion peaks that are present in the unassisted gait of unilateral SCI subjects. On the contrary, the simulated hip kinematics do not reveal side-specific features despite a 50\% imbalance in the iliopsoas strength of one of the subjects (2)(table \ref{tab:subject_muscles}). Interestingly, the simulation for subject 1 did not converge in the non-assisted scenario. We believe this was caused by insufficient actuation strength from the muscles. Similarly to the bilateral case, the RMSE for the non-AFO sides is smaller than for the AFO sides.

For the assisted simulations the model can portray the differences in ROM between the left and right ankle caused by the AFO, predicting the asymmetric characteristics observed in the individual's data. On the other side, concerning the knee and hip, the model produces similar kinematics for the swing phase but not during the stance phase. For these joints, the model converges to a symmetric movement which does not correspond to the observed data. The RMSE values, as in the case of the bilateral simulations similar to the ones obtained in previous studies\cite{Waterval2021ValidationGait}

\begin{table*}[htbp]
\centering
\caption{RMSE values for the moments of the gait simulations with and without AFO use.}
\label{tab:moments}
\begin{tabular}{@{}llllllllllllll@{}}
\toprule
\multicolumn{4}{l}{\textit{\textbf{Unilateral Case}}} & \multicolumn{3}{l}{RMSE} &  & \multicolumn{4}{l}{\textit{\textbf{Bilateral Case}}} & \multicolumn{2}{l}{RMSE} \\ \midrule
\multicolumn{4}{l}{Subject   ID} & 1 & 2 & 3 &  & \multicolumn{4}{l}{Subject ID} & 4 & 5 \\
\multirow{6}{*}{\begin{tabular}[c]{@{}l@{}}AFO\\ Supported\\ Gait\end{tabular}} & \multirow{3}{*}{\begin{tabular}[c]{@{}l@{}}AFO\\ side\end{tabular}} & \multirow{3}{*}{\begin{tabular}[c]{@{}l@{}}Moments\\ {[}Nm/kg{]}\end{tabular}} & Ankle & 0.96 & 3.31 & 1.21 &  & \multirow{6}{*}{\begin{tabular}[c]{@{}l@{}}AFO\\ Supported\\ Gait\end{tabular}} & \multirow{3}{*}{\begin{tabular}[c]{@{}l@{}}AFO side\\ (R)\end{tabular}} & \multirow{3}{*}{\begin{tabular}[c]{@{}l@{}}Moments\\ {[}Nm/kg{]}\end{tabular}} & Ankle & 0.89 & 3.34 \\
 &  &  & Knee & 0.91 & 4.02 & 1.77 &  &  &  &  & Knee & 6.81 & 2.18 \\
 &  &  & Hip & 1.37 & 6.47 & 1.07 &  &  &  &  & Hip & 2.64 & 1.99 \\
 & \multirow{3}{*}{\begin{tabular}[c]{@{}l@{}}Non AFO\\ side\end{tabular}} & \multirow{3}{*}{\begin{tabular}[c]{@{}l@{}}Moments\\ {[}Nm/kg{]}\end{tabular}} & Ankle & 1.22 & 2.65 & 1.35 &  &  & \multirow{3}{*}{\begin{tabular}[c]{@{}l@{}}AFO side\\ (L)\end{tabular}} & \multirow{3}{*}{\begin{tabular}[c]{@{}l@{}}Moments\\ {[}Nm/kg{]}\end{tabular}} & Ankle & 0.98 & 1.92 \\
 &  &  & Knee & 1.68 & 3.88 & 1.56 &  &  &  &  & Knee & 3.24 & 1.71 \\
 &  &  & Hip & 1.08 & 4.09 & 0.84 &  &  &  &  & Hip & 1.92 & 1.09 \\
 &  &  &  &  &  &  &  &  &  &  &  &  &  \\
\multirow{6}{*}{\begin{tabular}[c]{@{}l@{}}Non\\ Supported\\ Gait\end{tabular}} & \multirow{3}{*}{\begin{tabular}[c]{@{}l@{}}AFO\\ side\end{tabular}} & \multirow{3}{*}{\begin{tabular}[c]{@{}l@{}}Moments\\ {[}Nm/kg{]}\end{tabular}} & Ankle & - & 1.66 & 1.40 &  & \multirow{6}{*}{\begin{tabular}[c]{@{}l@{}}Non\\ Supported\\ Gait\end{tabular}} & \multirow{3}{*}{\begin{tabular}[c]{@{}l@{}}AFO side\\ (R)\end{tabular}} & \multirow{3}{*}{\begin{tabular}[c]{@{}l@{}}Moments\\ {[}Nm/kg{]}\end{tabular}} & Ankle & 1.43 & 1.07 \\
 &  &  & Knee & - & 8.17 & 3.46 &  &  &  &  & Knee & 13.51 & 1.96 \\
 &  &  & Hip & - & 6.59 & 1.28 &  &  &  &  & Hip & 3.16 & 1.20 \\
 & \multirow{3}{*}{\begin{tabular}[c]{@{}l@{}}Non AFO\\ side\end{tabular}} & \multirow{3}{*}{\begin{tabular}[c]{@{}l@{}}Moments\\ {[}Nm/kg{]}\end{tabular}} & Ankle & - & 0.97 & 1.22 &  &  & \multirow{3}{*}{\begin{tabular}[c]{@{}l@{}}AFO side\\ (L)\end{tabular}} & \multirow{3}{*}{\begin{tabular}[c]{@{}l@{}}Moments\\ {[}Nm/kg{]}\end{tabular}} & Ankle & 0.81 & 0.88 \\
 &  &  & Knee & - & 7.40 & 2.02 &  &  &  &  & Knee & 5.64 & 1.52 \\
 &  &  & Hip & - & 7.21 & 1.10 &  &  &  &  & Hip & 3.16 & 0.99 \\ \cmidrule(lr){4-8} \cmidrule(l){12-14} 
\end{tabular}
\end{table*}

\subsection{Moments}
Regarding the moments, the RMSE values seem to be correlated with the muscular state of the subjects being inversely proportional to the general muscular strength as seen in table \ref{tab:moments}. Not only in the bilateral but also in the unilateral case, the subjects with lower scores in the manual muscle testing scale (2 and 4) considerably present higher RMSE. As previously mentioned, the gait simulations for subject 1, who has a very poor muscular state, did not converge for the unassisted case. This model only generated stable gait in the presence of AFO torque input (AFO-assisted gait) even though the corresponding subject was able to walk without any orthotic device. We consider this could be due to the strong influence of dynamics in the cost function, as it considers the walk energy cost, the muscle activation and head acceleration\cite{Veerkamp2021EvaluatingGait,Waterval2021ValidationGait} which are dynamic related variables poorly represented in a model with considerable less strong actuators.

\section{Discussion}
The developed subject-adjusted musculoskeletal models for AFO-assisted and unassisted spinal cord injury gait partially capture the unassisted kinematics of hip and knee but still present difficulties in recreating the ankle kinematics, besides being previously used for bilateral calf weakness, which also presents debilitated ankle flexors. Also, the model can show the imbalances in the muscular states of the subjects and exhibit the effects of an ankle foot orthosis in the range of motion for the ankle and knee joint. 

In the same line of other studies focused on the simulation of pathological gait, there are still significant differences between the predicted and real SCI AFO kinematics with a standard deviation-normalized RMSE between one and two SD, as in the case of the simulation of spasticity\cite{Veerkamp2023PredictingHyperreflexia}. Also, the mean error for the angles in some gait phases can be observed to be higher than 10 degrees which is also observed in the simulation of healthy gait and cerebral palsy gait\cite{Park2023BidirectionalConditions,VanDenBosch2022BuildingStudy}.

We hypothesize that the model requires a higher number of actuators, and the addition of coronal movement as the current set of 9 muscles per limb might be unable to represent the complexity of SCI gait, Which has several neuronal and muscular contributions. This was evidenced in the case of the unassisted simulation of subject 1 where the current pipeline was unable to converge the model into a stable gait.

The developed models performed better in estimating kinematics than moments as evidenced in the RMSE values. In addition, the cost function which considers the walk energy cost, the muscle activation and head acceleration\cite{Veerkamp2021EvaluatingGait,Waterval2021ValidationGait} might not be suitable to generate SCI gait dynamics within this model. 

We believe that adding optimization of subject-specific dynamic, kinematic and muscular terms not represented in the current models is required to achieve functional outcomes in the simulations. 
Moreover, with sufficient data a close-loop supervised method could be used to achieve a complete parametrized model as proposed in  \cite{Krechowicz2023MachineSensors, Pal2022PersonalizedLearning}.

\section{CONCLUSIONS}
We proposed and evaluated a musculoskeletal model to simulate AFO-assisted and unassisted gait for SCI individuals wearing unilateral and bilateral AFO, in this preliminary study of five subjects we generated a first attempt to simulate the effects of orthotic aids for pathological gait.
One particularly interesting point in simulation is the difficulty in asymmetric gait. Currently, the model considers the asymmetries in the muscular states of the subject's ankle flexors as seen in the differential range of motion in the ankle joint in the non-assisted simulations while generating more symmetrical knee and hip kinematics. 
Similarly. when predicting the effect of the AFO on the gait of SCI subjects, the simulated kinematics reflect changes mostly in the range of motion. 
Even though the kinematic profile is somehow preserved for the knee and hip joints there are considerable differences in the ankle joint. 
We believe that the biomechanics of spinal cord injury are highly subject-specific, thus, the model also requires a higher dimensional parametrization. This could be achieved through additional actuators or by introducing subject-specific parameters that should be optimized in a closed-loop manner to convey more appropriate results.

\footnotesize

\bibliographystyle{unsrt}
\bibliography{references}

\normalsize
\end{document}